\documentclass{ws-ijmpcs}

\begin{document}

\def\be{\begin{eqnarray}}
\def\ee{\end{eqnarray}}

\markboth{C.G. Beneventano and E.M. Santangelo}
{Boundary conditions in the Dirac approach to graphene}

\catchline{}{}{}{}{}
%
%%%%%%%%%%%%%%%%%%%%%%%%%%%%%%%%%%%%%%%%%%%%%%%%%%%%%%%%%%%%%%%%%%%%

\title{\textbf{BOUNDARY CONDITIONS IN THE DIRAC APPROACH TO GRAPHENE DEVICES}}

\author{C.G. BENEVENTANO\footnote{Member of CONICET}}

\address{\textit{Departamento de F\'isica , Facultad de Ciencias Exactas, Departamento de Ciencias B\'asicas, Facultad de Ingenier\'ia, Universidad Nacional de La Plata and Instituto de F\'isica de La Plata, CONICET, calle 49 y 115\\ La Plata, 1900, Argentina}\\
gabriela@fisica.unlp.edu.ar}

\author{E.M. SANTANGELO\footnote{Member of CONICET}}

\address{\textit{Departamento de F\'isica , Facultad de Ciencias Exactas, Universidad Nacional de La Plata and Instituto de F\'isica de La Plata, CONICET, calle 49 y 115\\ La Plata, 1900, Argentina}\\
mariel@fisica.unlp.edu.ar}

\maketitle

\begin{history}
\received{Day Month Year}
\revised{Day Month Year}
\end{history}

\begin{abstract}
We study a family of local boundary conditions for the Dirac problem corresponding to the continuum limit of graphene, both for nanoribbons and nanodots. We show that, among the members of such family, MIT bag boundary conditions are the ones which are in closest agreement with available experiments. For nanotubes of arbitrary chirality satisfying these last boundary conditions, we evaluate the Casimir energy via zeta function regularization, in such a way that the limit of nanoribbons is clearly determined.

\end{abstract}

\ccode{PACS numbers: 72.80.Vp, 03.70.+k}

\section{\textbf{Introduction}}
\label{intro}

It would be redundant to start this paper with a detailed description of the wonderful properties of graphene. In this respect, the interested reader is referred to Ref.~\refcite{geim}.
As predicted theoretically\cite{semenoff,mele} twenty years before its production in a laboratory,\cite{novo1} electron transport in graphene is described by a massless Dirac equation, which leads to distinctive electronic properties.\cite{banda,ando} Indeed, many such properties were studied experimentally, starting with the determination of the quantum Hall effect,\cite{novo2} and found to agree with the predictions of a ``relativistic'' and massless Dirac field theory.\cite{gus}

The aforementioned properties allow to envisage many possible applications. However, a crucial point in achieving such goals as the construction of graphene-based transistors is the opening of a controllable band gap in an otherwise gapless material. The use of samples of finite size is a natural guess when trying to do so.\cite{ando} In fact, several measurements of the electric conductivity in graphene devices do show the existence of a gap.\cite{kim1,lin,molitor,kim2}

Unlike the case of usual semiconductors, the confinement of charge carriers to a finite region cannot be modeled, in the continuous Dirac theory, by the condition that the fields vanish at the boundaries. Most theoretical approaches to this problem presuppose an orientation dependence of the adequate boundary conditions,\cite{brey,akh} which is in contradiction with the experimental results.\cite{kim1} A nice general study of possible boundary conditions in the Dirac problem and of their symmetries in the case of nanotubes was presented in Ref.~\refcite{falko}.

The aim of this paper is twofold: in the first place, we will study a family of boundary conditions satisfying rather standard criteria, and compare the predictions arising from taking different members of such family with experimental results. After having shown that, among these members, the ones corresponding to MIT bag boundary conditions\cite{mit} are good candidates to reproduce the experimental outcomes, we will evaluate the Casimir energy\cite{casimir,michael} for nanotubes of arbitrary chirality satisfying such boundary conditions, and arrive to an expression which leads to a direct identification of the Casimir energy in the nanoribbon limit, and shows clearly the (exponentially decreasing) corrections due to finite length-to-width ratio.

\section{\textbf{General setting. Edge states}}
\label{edge}
We will choose the orientation of the lattice as in Ref.~\refcite{semenoff}, so that, by taking the two nonequivalent Dirac points as $K_{\pm}=(0,\pm \frac{4\pi}{3\sqrt{3}a})$, we get the total Hamiltonian as a direct sum of
\be H_{\pm}=\hbar v_F (-i\sigma_2\partial_x \pm i\sigma_1\partial_y)\,,\label{h}\ee
where $v_F=\frac{3at}{2\hbar}$ is the Fermi velocity of graphene, with $a=0.14nm$ the distance between nearest neighbors and $t=2.7eV$ the nearest neighbor hopping energy.

Such Hamiltonian corresponds to a free Dirac equation in $2+1$ dimensions, where the gamma matrices are given, in each valley, by $\gamma^0_{\pm}=i\sigma_3$, $\gamma^1_{\pm}=\sigma_1$, $\gamma^2_{\pm}=\pm\sigma_2$. We will study the corresponding eigenvalue problems $H_{\pm}\Psi_{\pm}(x,y)=E_{\pm}\Psi_{\pm}(x,y)$, when the domain of the differential operator is defined by a family of local boundary conditions which

\begin{romanlist}[(iii)]
\item{Are separately imposed in each valley,}
\item{Give a vanishing flux of current perpendicular to the boundary,\cite{akh}}
\item{Are defined through a self-adjoint projector.\cite{akh,falko}}
\end{romanlist}

Note that the condition (ii) is as close to confinement as one can get in a Dirac theory and it leads to a self-adjoint Hamiltonian (thus, to real energies).

From now on, we will study the problem around $K_{+}$, leaving the discussion on how to combine boundary conditions in both valleys for later on. We will consider the boundary to be placed at a given $x=x_0$ value. Throughout our calculations, we will take $v_F=\hbar =1$, and recover the right units when comparing our predictions with experimental results.

We start our analysis with the case of the half plane, with its boundary at $x=0$. For such geometry, it is easy to check that the current perpendicular to the boundary is proportional to $\Psi_{+}^{\dag}\sigma_2\Psi_{+}$, while the current along the boundary is proportional to $\Psi_{+}^{\dag}\sigma_1\Psi_{+}$. So, the most general local boundary conditions satisfying conditions (i) to (iii) above are given by $(I+\sigma_1\,e^{-i{\alpha}\sigma_2})\Psi_{+}\rfloor_ {x=x_0}=0$, which is a one-parameter family. In Ref.~\refcite{akh}, this family was associated to the existence of a staggered potential, different for $A$ and $B$ sites, in a total of $2N$ rows closest to a zigzag edge parallel to the $y$ axis, as in our case.

Note that $\alpha=0,\pi$ correspond to the so-called MIT bag boundary conditions,\cite{mit} while $\alpha=\pm \frac{\pi}{2}$ are the conditions used to mimic a zigzag boundary.\cite{brey} Here, we point out that, among the graphene community, MIT boundary conditions are usually called infinite-mass or Berry-Mondrag\'on boundary conditions because these two authors studied them in the $2+1$ context of quantum billiards, in Ref.~\refcite{berry}.

Each member of this family of boundary conditions imposes a different constraint on the density of tangential current at the boundary. In fact, one has $\Psi_{+}^{\dag}\sigma_1\Psi_{+}\rfloor_{x=x_0}=-\cos{(\alpha)}\Psi_{+}^{\dag}\Psi_{+}\rfloor_{x=x_0}$. In particular, zigzag boundary conditions enforce the tangential current to vanish at the boundary, while MIT ones equate it to the density of charge.

In view of the translational invariance along the $y$ direction we will propose, for each $k_{y}$, $\Psi_{+}(x,y)=e^{ik_y\,y}\psi_{+}(x)$. In order to analyze the existence of edge states (or the lack thereof), it is convenient to perform a unitary transformation of the eigenfunctions, $\tilde{\psi}(x)=e^{-i{\frac{\alpha}{2}\sigma_2}}\psi_{+}(x)$. This leads us to the eigenvalue problem,
\be
&&\left[-i\sigma_2\partial_x -\sigma_1k_y \cos{\alpha}+\sigma_3k_y \sin{\alpha}\right]\tilde{\psi}(x)=E\tilde{\psi}(x) \nonumber \\
&&\left(I+\sigma_1\right)\tilde{\psi}(x=0)=0\,,
\ee
together with the normalizability condition when $x\rightarrow \infty$.

It is a simple exercise to show that, for all $\alpha\neq  0,\pi$ (i.e., all boundary conditions different from the MIT bag ones) there are, apart from bulk sates, some edge states, corresponding to $E=k_y \cos{\alpha}$, with $k_y \sin{\alpha}> 0$, which are eigenfunctions decreasing exponentially with $x$, thus concentrated at the boundary.

The existence of these states is well known for zigzag boundary conditions ($\alpha=\pm \frac{\pi}{2}$), in which case they are zero energy modes.\cite{banda} This shows that, in a compact region with a single smooth boundary, these two boundary problems will not satisfy the Lopatinsky-Shapiro condition (equivalently, they will not define a Fredholm operator).\cite{elipticidad} In the remaining cases, the edge states correspond to real, nonzero, energies. We will comment on this point when treating circular quantum dots, in section \ref{dots}.

\section{\textbf{Graphene nanoribbons}}
\label{ribbons}

For a comparison with experiments, we will first discuss the case of a graphene nanoribbon, which requires the imposition of a boundary condition at a second boundary, placed at $x=W$.

Experiments concerning nanoribbons\cite{kim1,lin,kim2} show a gap which, moreover, is symmetric around zero gate voltage. There are two ways of obtaining a symmetric spectrum, i.e., choosing exactly the same projector to define the boundary condition at $x=W$ or choosing, instead, the orthogonal one. It is easily shown that the first alternative allows for the existence of zero modes, no matter the value of $\alpha$. They appear for all values of $k_y$ when $\alpha=\pm  \frac{\pi}{2}$ and for $k_y =0$ in the remaining cases. So, we will limit our discussion to the second alternative, which is also consistent with the fact that the sign of the inward normal is opposite at both boundaries.

Our boundary value problem will now be,
\be
&&\left[-i\sigma_2\partial_x -\sigma_1k_y \cos{\alpha}+\sigma_3k_y \sin{\alpha}\right]\tilde{\psi}(x)=E\tilde{\psi}(x) \nonumber \\
&&\left(I+\sigma_1\right)\tilde{\psi}(x=0)=0\nonumber \\&&\left(I-\sigma_1\right)\tilde{\psi}(x=W)=0\,.
\label{problem}\ee

In all cases one has $E=\pm \sqrt{k_x^2 +k_y^2}$. However, MIT bag boundary conditions are unique in that the spectrum is determined by the equation $\cos{(k_x W)}=0$, which does not depend on $k_y$, and they allow only real values of $k_x$. Thus we have, in these two cases ($\alpha=0,\pi$),
\be
E_n=\pm \sqrt{\left(\frac{(n+\frac12)\pi}{W}\right)^2+k_y^2}\quad n=0,...,\infty\,.
\label{espectrobag}\ee

This leads to an energy gap $\Delta_E = \frac{\pi}{W}$. We stress that the same dispersion formula was obtained from a microscopic model in Ref.~\refcite{onipko}.

The remaining values of $\alpha$, instead, lead to a spectrum determined by
\be
k_x \cos{(k_x W)}=k_y \sin{\alpha}\sin{(k_x W)}\quad {\rm for}\, E\neq \pm k_y\,,
\label{espectro1alfa}\ee
and
\be
k_y=\frac{1}{W\sin{\alpha}},\quad {\rm for}\, E= \pm k_y\,.
\label{espectro2alfa}\ee

Note that both equations break the invariance under $k_y \rightarrow -k_y$. It is difficult to imagine why this invariance would be broken in a ribbon, which extends to $-\infty <y<\infty$, in the absence of electromagnetic fields. Such invariance could only be recovered by imposing exactly the same boundary conditions on the eigenfunctions around the other valley.

Moreover, for $k_y=0$ one has $k_x=\frac{(n+\frac12)\pi}{W}$, no matter the value of $\alpha$. For all the remaining values of $k_y$, at variance with the situation in the MIT case, the admissible values of $k_x$ are not equally spaced.

But, more important, Eq. (\ref{espectro1alfa}) allows for imaginary as well as real values of $k_x$. Calling $\kappa =i\,k_x$ one has, for $E\neq \pm k_y $,
\be
\kappa \cosh{(\kappa W)}=k_y \sin{\alpha}\sinh{\!(\kappa W)}, \,{\rm for}\,|k_y |\!>\!\frac{1}{W|\sin{\alpha}|}\,\,.
\label{espectro3alfa}
\ee

When $|\sin{\alpha}|=1$, i.e., zigzag boundary conditions, this equation allows for energies arbitrarily close to zero when $\kappa\rightarrow \infty$. As a consequence, no gap exists in this case, which is a well known fact.\cite{banda} For the remaining values of $\alpha$, the eigenenergies coming from Eqs. (\ref{espectro1alfa}) and (\ref{espectro3alfa}) never tend to zero. The analysis of the minimal value of $|E|$ can be performed analytically. From such analysis one concludes that, for all $\alpha$, the energy gap satisfies $\Delta E \leq \frac{\pi}{W}$.

The experiments on nanoribbons\cite{kim1,lin,kim2} show a transport gap as a function of the gate voltage, when performed at low temperature and bias voltage. This eliminates zigzag boundary conditions as candidates to describe the physical situation. For the remaining values of $\alpha$ we have, recovering units, $\Delta E \leq \frac{\hbar v_F\pi}{W}=\frac32 \pi t\frac{a}{W}$. The equal sign holds only for MIT bag boundary conditions ($\alpha=0,\pi$).

As for the numerical value of the gap, Ref.~\refcite{kim2} presents a study of several graphene nanoribbons of different widths, all of which show a gap in the gate voltage corresponding to a one-particle energy gap fitted to $\Delta_m =36 eV \frac{a}{W}$. This is roughly three times our result for MIT boundary conditions, i.e., $\Delta E =12.7 eV \frac{a}{W}$. Ref.~\refcite{lin} finds, for a sample of width $W=30nm$, a value of the energy gap $\Delta E =46 meV$, in much better agreement with our result for such case, i.e., $\Delta E =57 meV$ (note, that, in this case, our prediction is higher than the measured gap). Obviously, both experiments disagree. The origin of such discrepancy is not clear to us, since both use similar values of the capacitance for comparable samples. So, numerical values of the gap cannot be used, at present, to select among different non-zigzag values of $\alpha$. However, Ref.~\refcite{lin} shows equally spaced plateaux in the conductivity, which only MIT bag boundary conditions can explain. Thus, these are the best candidates for the boundary conditions to be imposed in the continuous model.

Finally, as shown clearly by Ref.~\refcite{kim1}, and at odds with most theoretical models\cite{brey,akh} (which would impose different boundary conditions depending on the orientation of the boundary) the measured gap in the gate voltage doesn't depend on the orientation of the boundary. This will certainly be the case in our continuous model if MIT bag boundary conditions are written as $(I+/\!\!\!n)\,\psi(x=0,W)=0$, where $n$ is the inward (or outward, which leads to the same result) pointing normal vector corresponding to each boundary. Moreover, consistently imposing MIT boundary conditions around $K_{-}$ leads to the same spectrum in both valleys.

\section{\textbf{Nanodots}}
\label{dots}

In order to further compare the predictions of different members of our family of boundary conditions with the experiment, we will now treat the case of a circular graphene dot of radius $R$. To this end, we adopt polar coordinates. Taking the gamma matrices for the theory around $K_{+}$ as before, we are led to the boundary value problem (with $v_F =\hbar =1$),
\be
&&\left[-i{\gamma}^{\theta}\partial_r +i\frac{{\gamma}^{r}}{r}{\partial}_{\theta}\right]\psi(r,\theta)=E\psi(r,\theta) \nonumber \\
&&\left(I-{\gamma}^{r}e^{-i\alpha {\gamma}^{\theta}}\right)\psi(r=R,\theta)=0\nonumber \\&&\psi(r,\theta)=\psi(r,\theta +2\pi)\,,
\label{problem2}\ee
where ${\gamma}^{r}=\sigma_1 \cos{\theta}+\sigma_2 \sin{\theta}$ and ${\gamma}^{\theta}=\sigma_2 \cos{\theta}-\sigma_1 \sin{\theta}$.

Solving this boundary value problem is a simple exercise (see, for instance, Ref.~\refcite{peres}). The first outcome is that zigzag boundary conditions ($\alpha=\pm \frac{\pi}{2}$) allow for an infinite amount of zero modes, as expected from the facts that they do not satisfy the Lopatinski-Shapiro condition and that we are now treating a compact region with a smooth boundary. As for the remaining conditions in the family, none of them allows for zero modes. Since experiments on quantum dots also present a gap, we will concentrate on the remaining cases ($\cos{\alpha}\neq 0$), which give a spectrum determined by
\be
(1\!-\!\sin{\alpha})J_{n}(|E|R)\!+\!\!s\cos{\alpha}J_{n+1}(|E|R)=0,\, n=0,...,\infty\nonumber \\
(1\!-\!\sin{\alpha})J_{n+1}(|E|R)\!-\!\!s\cos{\alpha}J_{n}(|E|R)=0,\, n=0,...,\infty
\label{espectrodot}\ee
where $J_{n}$ is the Bessel function of order $n$, and $s$ is the sign of the energy.

Now, the experiment\cite{molitor} shows a gap in a quantum dot which is symmetric around zero gate voltage. This, again, points to the MIT boundary conditions as good conditions to impose on the continuum model in order to reproduce the experimental results, since all the remaining values of $\alpha$ produce a spectral asymmetry, as can be easily seen from Eqs. (\ref{espectrodot}).

\section{\textbf{Casimir energy of graphene nanotubes and nanoribbons}}
\label{Casimir}

Now that we have justified the use of MIT bag boundary conditions when treating graphene devices in the continuum limit, particularly in the case of ribbons, it's interesting to calculate the vacuum expectation value of the energy, also known as Casimir energy, due to the finite size of a given sample. To achieve this goal, we will evaluate such magnitude, in the framework of the zeta function regularization, for a general graphene nanotube of arbitrary chirality (characterized by $\frac{\delta}{2}$), compactification length $L$ and finite length $W$ in the perpendicular direction, imposing MIT bag boundary conditions at the boundaries $x=0,W$. The limit corresponding to graphene nanoribbons is $\frac{L}{W}\rightarrow \infty $. So, we will analytically extend the otherwise diverging sum of allowed energies in such a way that this limit is made patent, and the exponential corrections due to finite length effects are clearly displayed. A similar calculation was already presented in Ref.~\refcite{ads}, in the particular case of an antiperiodic compactification ($\frac {\delta}{2}=\frac12$). Here, we generalize the calculation to arbitrary $0\leq \frac{\delta}{2}<\frac12$, which includes the interesting cases of conducting nanotubes ($\frac {\delta}{2}=0$) and semiconducting ones ($\frac {\delta}{2}=\frac13$). Due to the invariance of the spectrum under $\frac {\delta}{2}\rightarrow \frac {\delta}{2}+1$ and $\frac {\delta}{2}\rightarrow 1-\frac {\delta}{2}$, this will cover all the possible twists in the compact direction.

The Casimir energy of the problem at hand, will be regularized according to
\be
\left. E_C =-\frac{g_s\,g_v}{2} \left[\sum_{E_{n,l}>0} E_{n,l}^{-s}+\sum_{E_{n,l}<0} |E_{n,l}|^{-s}\right]\right\rfloor_{s=-1}\,,
\ee
where $E_{n,l}$ are the modes in Eq. (\ref{espectrobag}), with $k_y=\frac{2\pi}{L}(l+\frac{\delta}{2})$, to account for the different possible chiralities of the nanotube, and $g_s=2$ is the spin degeneration and $g_v=2$ is the valley degeneration.

More explicitly, for the Casimir energy per unit compactification length, we have
\be
\frac{E_{C}}{L}=-\frac{g_s\,g_v}{L} \sum_{l=-\infty}^{\infty}\sum_{n=0}^{\infty}\left.\left[\left[(n+\frac{1}{2})\frac{\pi}{W}\right]^2+\left[(l+\frac{\delta}{2} )\frac{2\pi}{L}\right]^2\right]^{-\frac{s}{2}}\right\rfloor_{s=-1}
\label{once}\ee
or, after Mellin transforming the previous expression,
\be
\left.\frac{E_{C}}{L}=-\left(\frac{2\pi}{L}\right)^{\!-s}\!\frac{g_s\,g_v}{L\Gamma\left(\frac{s}{2}\right)}\int_0 ^{\infty}\!\!dt\,t^{\frac{s}{2}-1}\!\!\sum_{l=-\infty}^{\infty}\sum_{n=0}^{\infty}\!e^{-t\left\{\left[(n+\frac{1}{2})\frac{L}{2W}\right]^2+
\left[l+\frac{\delta}{2}\right]^2\right\}}\!\right\rfloor_{s=-1}\!.
\label{e22}\ee
An analytical extension, convenient to isolate the term remaining in the nanoribbon limit, can be performed by making use of the inversion formula\cite{klaus}
\be
\sum_{l=-\infty}^{\infty}e^{-t(l+c)^2}=\left(\frac{\pi}{t}\right)^{\frac12}\sum_{l=-\infty}^{\infty}
e^{-\frac{{\pi}^2\,l^2}{t}-2\pi i\, l c}\,.
\label{inver}\ee
When doing so, we get
\be
\nonumber\frac{E_{C}}{L}=\!-\!\left(\frac{2\pi}{L}\right)^{\!\!\!-s}\!\frac{g_s g_v {\pi}^\frac12}{L\Gamma\left(\frac{s}{2}\right)}
\left\{4\sum_{l=1}^{\infty}\sum_{n=0}^{\infty}\!\cos{(\pi l \delta)}\!\int_0 ^{\infty}\!\!\!dt\,t^{\frac{s-1}{2}-1}e^{-t\left[(n+\frac12)\frac{L}{2W}\right]^2-\frac{{\pi}^2 l^2}{t}} \right.\\ + \left.\left.\sum_{n=0}^{\infty}\int_0 ^{\infty}dt\,t^{\frac{s-1}{2}-1}e^{-t\left[(n+\frac12)\frac{L}{2W}\right]^2}\right\}\!\right\rfloor_{s=-1}.
\ee

Now, after performing the integral in the first term, and writing the second one as a Hurwitz zeta function ($\zeta_H$), we obtain
\be
\nonumber \frac{E_{C}}{L}\!=\!-\!\left(\frac{2\pi}{L}\right)^{\!\!\!\!-s}\!\!\frac{g_s g_v {\pi}^\frac12}{L\Gamma\left(\frac{s}{2}\right)}
\!\left\{\!4\!\sum_{l=1}^{\infty}\sum_{n=0}^{\infty}\!\cos{(\pi l \delta)} \!\left(\!\!\frac{\pi\,l}{\left(n+\!\frac12\!\right)\frac{L}{2W} }\!\!\right)^{\!\!\frac{s-1}{2}}\!\!\!\!\!\!\!K_{\frac{s-1}{2}}\!\!\left(\!\frac{\left(n+\!\frac12\right)l\pi L}{W}\!\right) \right.\\ \!\!\!\!\!\!\!\!\! +  \left.\left.\Gamma\left(\frac{s-1}{2}\right)(\frac{L}{2W})^{1-s}\zeta_H (s-1,\frac12)\right\}\right\rfloor_{s=-1}\!\!.
\ee

Finally, after relating the Hurwitz zeta function to the corresponding Riemann one ($\zeta_R$), and using the reflection formula for this last,\cite{gradshteyn} we obtain the final expression for the Casimir energy per unit compactification length, i.e.,
\be
\frac{E_{C}}{L}=\frac{2g_s\,g_v}{LW}\!\!\sum_{l=1,n=0}^{\infty}\!\!\cos{(\pi\,l\,\delta)}\frac{\left(n+\frac12\right) }{l}\,K_1 \!\left(\!\frac{\left(n+\frac12\right)\pi\,l\, L}{W}\right)-\frac{3g_s\,g_v}{32\pi{W}^2}\,\zeta_R (3)\,.
\label{ec1}\ee

Not unexpectedly, the nanoribbon limit ($\frac{L}{W}\rightarrow \infty$) of this Casimir energy, given by $\frac{E_{C}}{L}=-\frac{3g_s\,g_v}{32{W}^2}\zeta_R (3)$, turns out to be independent of the chirality of the nanotube. Indeed, it could have been obtained, by considering a continuous $k_y$ in Eq. (\ref{once}), as
\be
\frac{E_{C}}{L}=-\frac{g_s\,g_v}{2\pi} \int_{-\infty}^{\infty}\,dk_y\,\sum_{n=0}^{\infty}\left.\left[\left[(n+\frac{1}{2})\frac{\pi}{W}\right]^2+k_y^2\right]^{-\frac{s}{2}}\right\rfloor_{s=-1}
\,.\ee

The numerical value of this Casimir energy for nanoribbons is, after recovering physical units,
\[\frac{E_{C}}{L}=-\frac{3\hbar v_F}{8\pi{W}^2}\zeta_R (3)=-0.1435\frac{\hbar v_F}{{W}^2}\,,\]
which corresponds to an attractive force per unit length $\frac{F_{C}}{L}=-0.2870\frac{\hbar v_F}{{W}^3}$.

Apart from reproducing this limit, Eq. (\ref{ec1}) gives the explicit (exponentially decreasing for $L>>W$) corrections to  this limit.

Here, it is interesting to note that an alternative expression for the Casimir energy of nanotubes was given in Eq. (69) of Ref.~\refcite{saharian}. Such expression, which is obtained through a combined use of the zeta function regularization and the generalized Chowla-Selberg formula, proves more useful in the limit $\frac{W}{L}\rightarrow \infty$. With our approach, such expression can be obtained by applying the inversion formula (\ref{inver}) to the sum over $n$ (instead of the sum over $l$ as we have just done) in Eq. (\ref{e22}). Thus one gets, for the Casimir energy per unit length of the nanotube

\be
\frac{E_{C}}{W}=\frac{2g_s\,g_v}{LW}\!\!\!\sum_{l=-\infty,n=1}^{\infty}\!\!\!\!\!(-1)^n \frac{|l+\frac{\delta}{2}|}{n}K_1 \!\!\left(|l+\frac{\delta}{2}|\frac{4n\pi\, W}{L}\right)+\frac{g_s\,g_v}{\pi{L}^2}\sum_{n=1}^{\infty}\frac{\cos{(n\pi\,\delta)}}{n^3}\,.
\label{tubo}\ee

This coincides with the result in Ref.~\refcite{saharian}, except for the fact that, in this reference, the spin degeneration hasn't been included, and there is a different sign (seemingly a misprint) in front of the double sum.

In the limit $\frac{W}{L}\rightarrow \infty$ (long nanotubes), one has $\frac{E_{C}}{W}=\frac{g_s\,g_v}{\pi{L}^2}\sum_{n=1}^{\infty}\frac{\cos{(n\pi\,\delta)}}{n^3}$. The corrections due to finite lenghts are also exponentially decreasing, except for the case of conducting nanotubes ($\delta =0$), where the $l=0$ term in the double sum is given by $-\frac{g_s\,g_v\pi}{6{W}^2}$.

\section{\textbf{Final comments and conclusions}}
\label{conclusions}

Even though the experiments on graphene devices are still at a very preliminary stage, we have shown in this paper that MIT bag boundary conditions are the most promising ones among the family of local boundary conditions that can reasonably (in the sense of our assumptions in section \ref{edge}) be imposed in the continuum limit. In particular, for nanoribbons, they predict not only the existence of a gap, which does not depend at all on the orientation, but also the existence of equally spaced energy levels. In the case of circular nanodots, they are the only ones, in the family of local boundary conditions studied here, predicting a spectrum which is symmetric around zero, as the experiment\cite{molitor} seems to imply.
Note that Ref.~\refcite{peres} had already suggested that this could be the case. In that reference, MIT bag boundary conditions are rather called Berry-Mondrag\'on boundary conditions, because they were also studied in Ref.~\refcite{berry}, in the context of quantum billiards. Indeed, this is the first guess a particle physicist would make when asked for confinement in a Dirac theory. Though experiments on graphene nanodevices are still at a very initial stage, the comparison with the experimental results presented in this paper shows this to be, most probably, the case. As for zigzag conditions, the fact that they fail to define a Fredholm Hamiltonian would be an obstacle to the formulation of a good quantum theory.

In view of the previous considerations, we have performed in this paper a full zeta function calculation of the Casimir energy for nanotubes of arbitrary dimensions and chiralities, which lead us to an expression particularly adequate to the case of nanoribbons. Such expression clearly shows that, in this limit, and independently of the chirality of the initial nanotube, there is an attractive force between the boundaries of the ribbon, and that the corrections to this force decay exponentially with $\frac{L}{W}$. For some previous work on Casimir energies in the presence of graphene sheets, we refer the interested reader to Refs.~\refcite{dima,ignat}.

\section*{Acknowledgments}

We thank the organizers of QFEXT11 for the warm hospitality and excellent working atmosphere enjoyed during the conference.

This work was partially supported by Universidad Nacional de La Plata (Proyecto 11/X492), CONICET (PIP 1787) and ANPCyT (PICT 909).

\end{document}